\documentclass[a4paper,11pt]{article}

\usepackage{lineno}
\usepackage{amssymb}
\usepackage{amsfonts,amsbsy,graphics,epsfig}
\usepackage{graphicx}
\usepackage{amsmath}
\usepackage{color}
\usepackage{multirow}
\usepackage{arydshln}

\usepackage{booktabs}
\usepackage{url}

\usepackage[dvipsnames]{xcolor}
\usepackage{wasysym}
\usepackage{natbib}
\usepackage{cleveref}
\usepackage{adjustbox}
\modulolinenumbers[1]

\usepackage{authblk}

\title{A functional-model-adjusted spatial scan statistic}

\author[]{Mohamed-Salem Ahmed}
\author[]{Micha\"el Genin}
\affil[]{\small {\textit{Univ. Lille, EA2694 - Sant\'e publique : \'epid\'emiologie et qualit\'e des soins, F-59000 Lille, France\\
mohamed-salem.ahmed@univ-lille.fr\\
michael.genin@univ-lille.fr}}}
\date{ }

\begin{document}
  \maketitle
  \begin{abstract}
This paper introduces a new spatial scan statistic designed to adjust cluster detection for longitudinal confounding factors indexed in space. The functional-model-adjusted statistic was developed using generalized functional linear models in which longitudinal confounding factors were considered to be functional covariates. A general framework was developed for application to various probability models. Application to a Poisson model showed that the new method is equivalent to a conventional spatial scan statistic that adjusts the underlying population for covariates. In a simulation study with univariate and multivariate models, we found that our new method adjusts the cluster detection procedure more accurately than other methods. Use of the new spatial scan statistic was illustrated by analysing data on premature mortality in France over the period from 1998 to 2013, with the quarterly unemployment rate as a longitudinal confounding factor. \\

\end{abstract}

\section{Introduction}\label{sec1}
In many fields of science, cluster detection methods are useful tools for objective identifying aggregations of events in time and/or space and for determining the latter's statistical significance. In the field of epidemiology, researchers often seek to detect spatial clusters in which the risk of disease is significantly higher or lower than in the rest of the geographical area studied. For diseases of unknown etiology, information on the presence and nature of clusters provides clues to the disease mechanism (especially in terms of environmental factors), and can facilitate the design of subsequent individual-level observational studies.

Over the last few decades, several cluster detection methods have been developed. In particular, spatial scan statistics (originally proposed by Kulldorff, on the basis of Bernoulli and Poisson models \citep[][]{kulldorff1997spatial,kulldorff1999spatial}) are powerful methods for detecting spatial clusters with a variable scanning window size and in the absence of pre-selection bias, and then testing the clusters' statistical significance. Following on from Kulldorff's initial work, several researchers have adapted spatial scan statistics to other spatial data distributions, such as ordinal \citep[][]{jung2007spatial}, normal \citep[][]{kulldorff2009scan}, exponential \citep[][]{huang2007spatial} and Weibull model \citep[][]{bhatt2014spatial}. Spatial scan statistics have been extended to the multivariate framework by \cite{kulldorff2007multivariate}, \cite{neill2012fast}, and, most recently, \cite{cucala2017multivariate, cucala2018multivariate}.

One of the main problems in cluster detection is the need to adjust for covariates. If a covariate is a confounding factor associated with the event of interest, and is not homogeneously distributed over a geographical area, a cluster analysis can generate clusters in which the covariate (and not the event of interest) predominates. For example, clusters of cardiovascular disease must be adjusted for social deprivation, which is a strong confounding factor \citep[][]{rothman2008modern}. In the absence of adjustment, the analysis may highlight very deprived areas that have a higher number of disease cases but are not epidemiologically relevant because of confusion bias. In the literature, several covariate adjustment techniques have been applied to spatial scan statistics. For the Poisson model, \cite{kulldorff1997breast} originally suggested the use of (i) indirect standardization methods to adjust for qualitative covariates, and (ii) regression methods to adjust for quantitative covariates and to estimate the expected number of cases per spatial unit. For the Bernoulli model, \cite{kulldorff2007multivariate} suggested using several datasets for each stratum of a qualitative covariate. \cite{klassen2005geographical} applied multilevel regression methods to adjust for quantitative covariates. More recently, Jung \citep[][]{jung2009generalized} used generalized multivariate linear models (GMLMs) to build spatial scan statistics that incorporated covariates. The latter approach is particularly valuable because it merges spatial scan statistics developed for different probability models into a single framework. However, this approach has limitations when dealing with longitudinal covariates. In a purely spatial analysis, there are two possible scenarios for longitudinal data: (i) the variable outcome and the covariates are observed on the same time scale (e.g. one observation per year for both) over a long period of time, or (ii) the variable outcome and the covariates are observed on different time scales (e.g. one observation per year for the outcome, and one observation per month for covariates). In Jung's approach, a simplistic way of managing longitudinal covariates in both scenarios is to summarize the data by averaging them (or determining the median) over the entire time period. However, this may lead to significant information loss and a decrease in the quality of covariate adjustment. Alternatively, the confounding factors for each measurement time scale can be included in the model as a covariate, as long as the time scale is the same for each of the spatial units (in order to limit the number of missing values). However, this approach may create a high-dimensional vector of coefficients and introduce multicollinearity \citep[][]{james2002generalized}.

In the present work, we developed a spatial scan statistic based on functional data analysis (FDA) \citep[][]{ramsay2005functional}. Firstly, our approach allows longitudinal data to be considered as the realization of a random function over an interval containing discrete time points. It should be noted that the random function can be observed at different, unequally spaced time points for each location. Secondly, our approach replaces the high-dimensional vector of coefficients by a parameter function to be estimated. These two characteristics make it possible to overcome both the above-mentioned problems, i.e. identical measurement times, and high dimensionality.

The present article is organized as follows. Section \ref{Methodology} describes the methodological aspects of the functional-model-adjusted spatial scan statistic (FMASSS). In Section \ref{Poisson}, the FMASSS was applied to a Poisson model, and was found to be equivalent to a conventional spatial scan statistic when the underlying population was adjusted for covariates. Section \ref{Simulation_study} presents both the design and the results of a simulation study. Section \ref{Real_data} describes the application of the FMASSS to epidemiologic data and the detection of clusters of high and low premature mortality in France. Lastly, the results are discussed in Section \ref{Discussion}.

\section{Functional-model-adjusted spatial scan statistic}
\label{Methodology}

Let consider that at each location $s_i$ (one of $n$ different spatial locations $s_1,\ldots,s_n$ included in $D\subset \mathbb{R}^2$)), we observe an outcome variable $Y_i$ and two type of covariate: $Z_i$ is a $p\times 1$ random vector and  $\{X_{i1},\ldots,X_{im_i}\}$ is the realization of a real-valued stochastic process at $m_i$ time points $t_{i1},\ldots,t_{im_i}$ (i.e. longitudinal data). Hereafter, all observations are considered to be independent, this is a classical assumption in scan statistics. A spatial scan statistic usually denotes the maximum concentration observed among a collection of potential clusters denoted by $\mathcal{S}=\{S_k \subset D,\, k=1,2,... \}$. It is used as a test statistic for areas in which the concentration might be abnormally high or abnormally low \cite{Cressie1977}. Kulldorff \cite{kulldorff1997spatial} introduced a spatial scan statistic based on a generalized likelihood ratio; this enables the comparison of concentrations in potential clusters of different sizes, and takes account of heterogeneity in the underlying population. Without loss of generality and in line with Kulldorff's work \citep[][]{kulldorff1997spatial}, we shall focus on variable-size, circular clusters. Hence, the set of potential clusters $\mathcal{S}$ is built so that (i) each potential cluster is centered at a particular location, and (ii) the radius is limited so that the corresponding cluster cannot cover more than 50\% of the studied region. It should be noted that many other configurations (such as elliptical clusters \citep[][]{kulldorff2006elliptic} and graph-based clusters \citep[][]{cucala2013spatial} have been suggested.

Conventionally, the spatial scan statistic can be defined as the potential cluster that maximizes a log-likelihood ratio (LLR) over $\mathcal{S}$ namely the most likely cluster (MLC). This LLR is based on a null hypothesis $\mathcal{H}_0$ (the absence of a cluster) and an alternative hypothesis $\mathcal{H}_1$ (the presence of a cluster). If confounding covariates ($Z_i$ and $\{X_{i1},\ldots,X_{im_i}\}$) are present, the MLC can be revealed by these factors alone. Thus, the spatial scan statistic has to be adjusted with respect to these covariates. In Jung's GMLM approach \citep[][]{jung2009generalized}, $Z_i$ and $\{X_{i1},\ldots,X_{im_i}\}$ will be integrated as separate covariates. However, as mentioned in the Introduction, this approach can be limited by information loss and high dimensionality. Hence, we developed an FMASSS that considers $\{X_{i1},\ldots,X_{im_i}\}$ as realizations of a random function $\{X_i(t),\, t \in \mathcal{T}\}$, where $\mathcal{T}$ is an interval containing the discrete time points. The random function $\{X_i(t),\, t \in \mathcal{T}\}$ is approximated from the longitudinal observations $\{X_{i1},\ldots,X_{im_i}\}$. More generally, a basis of functions $\{\varphi_j(t), \, j = 1,\ldots,K, \; t \in \mathcal{T}\}$ is considered with $K\leq \min(m_1,\ldots, m_n )$, and the random function is assumed to belong to the space generated by this basis 
\begin{equation}
X_i(t)=\sum_{j=1}^{K}a_{ij}\varphi_j(t),
\label{Xbasis}
\end{equation}
where the $n\times K$ matrix basis coefficients $A$ with elements $a_{ij}$ can be estimated using either an interpolation method (if the measurements $\{X_{i1},\ldots,X_{im_i}\}$ are observed without error, i.e $X_{ik}=X_i(t_{ik}),\, k=1,\ldots,m_{i}$, or an ordinary (or penalized) least-square method (if the measurements are observed with some error, i.e $X_{ik}=X_i(t_{ik})+e_{ik},\, k=1,\ldots,m_{i}$. The choice of the basis of functions depends on the shape of the longitudinal data. For instance, a B-spline basis is the most suitable choice for non-periodic functional data, a Fourier basis can be useful for periodic functional data, while a wavelet basis can be appropriate for functional data with discontinuities or changes in behavior (see \cite{ramsay2005functional} for more details).

Once the random function $\{X_i(t),\, t \in \mathcal{T}\}$ has been built for each location si, one can use the generalized functional linear model\cite{muller2005generalized} to adjust the spatial scan statistic with respect to the covariate $Z_i$ and the random function $\{X_i(t),\, t \in \mathcal{T}\}$. To this end, let $S_k\in \mathcal{S}$ and assume that the conditional mean of the outcome variable $Y_i$, (with respect to the covariate information and the potential cluster) is defined by the following revised generalized functional linear model:
\begin{equation}
\mu_i^{(k)}=E\left(Y_i\left\vert S_k,Z_i,X_i\right.\right)=\Phi^{-1}\left(\alpha+\delta_{k}\xi_{i}^{(k)}+Z_i^{'}\beta+\int_{\mathcal{T}} X_i(t)\theta(t)dt\right),
\label{GLFM}
\end{equation}
where $\xi_{i}^{(k)}$ is a binary covariate equal to 1 if the location $s_i$ belongs to $S_k$  and equal to 0 otherwise, and where $\Phi(\cdot)$ is a known increasing link function. The parameters of interest are the intercept $\alpha$, $\delta_k$ which refers to the intensity of the cluster, the coefficients $\beta$ associated with the $p \times 1$ vector of covariates $Z$, and the parameter function $\theta(\cdot)$, which is a smoothing function that can be considered as a generalization of a slope function.The parameters $\beta$ and $\theta(\cdot)$ are fixed inside and outside the potential cluster, which means that the distributions of the covariates $Z$  and $X(\cdot)$ are invariant with respect to the clustering hypotheses. In other words, the conditional mean of $Y_i$ inside $S_k$ is fully characterized by its intensity $\delta_k$. It should be noted that $\exp(\delta_k)$ can be interpreted as the covariate-adjusted relative risk for individuals within the potential cluster $S_k$, relative to the risk for those outside it. The clustering hypotheses can therefore be expressed as follows: 
\begin{equation*}
\left\{
\begin{array}{lcl}
\mathcal{H}_0 & : &   \delta_{k}=0\\ 
& & \\ 
\mathcal{H}_1 &: & \delta_{k}>0\quad ( \mbox{  or  } \delta_{k}<0).
\end{array}
\right.
\end{equation*}
Given that  $\Phi(\cdot)$ is an increasing function, $\mathcal{H}_1$ means that the mean of $Y_i$ inside $S_k$ is higher (or lower) than the mean of $Y_i$ outside $S_k$.

As mentioned above, the spatial scan statistic is based on the likelihood ratio between these two hypotheses. Thus, in order to provide a general framework that can handle various models (Bernoulli, normal, Poisson, etc.), one needs to assume that the outcome variable $Y$ has a known, parametrized, conditional log-likelihood function:
\begin{equation}
F\left(Y_i;\, \mu_{i}^{(k)}, \sigma\left(\mu_{i}^{(k)}\right)\right),
\label{LL}
\end{equation} 
where $\sigma(\cdot)$ is a positive function defining the variance of $Y$. \\

Below, we describe the estimation procedure under each hypothesis and then introduce the FMASSS.\\

\noindent {\it\textbf{Estimation under the null hypothesis}}. Under the null hypothesis, model (\ref{GLFM}) is reduced to a GLFM:
\begin{equation}
\mu_i=E\left(Y_i\left\vert Z_i,X_i\right.\right)=\Phi^{-1}\left(\alpha+Z_i^{'}\beta+\int_{\mathcal{T}} X_i(t)\theta(t)dt\right).
\label{muH0}
\end{equation}
We used the popular estimation procedure developed by M{\"u}ller and Stadtm{\"u}ller\cite{muller2005generalized}. It is based on a truncation strategy in which the random function and the parameter function are projected into a space of functions generated by a basis of functions with an arbitrary dimension. Let $\{\phi_j(t),\, j=1,\ldots,K\}$ be the eigenbasis associated with the functional principal component analysis (PCA) of the functional data $\{X_i(t),\, i=1,\ldots,n\}$. For a fixed $J$, the parameter function is approximated by its projection in the space of functions generated by the first $J$ eigenfunctions:
\begin{equation*}
\tilde{\theta}(t)=\sum_{j=1}^{J}\theta_j\phi_j(t).
\end{equation*}
Using this approach, \cite{muller2005generalized} suggested that the conditional mean (\ref{muH0}) could be approximated by its truncated version $\tilde{\mu}_{i}$:
\begin{equation}
\tilde{\mu}_i=\Phi^{-1}\left(\alpha+Z_i^{'}\beta+C_i^{'}\theta\right)
\label{TCmu}
\end{equation}
where $\theta=(\theta_{1},\ldots,\theta_{J})^{'}$ and $C_i$ is the coefficient vector of the random function $\{X_i(t),\, t \in \mathcal{T}\}$ in the eigenbasis, which is given by:
$$C_{ij}=\int_{T}X_{i}(t)\phi_{j}(t), \qquad j=1,\ldots,K.$$
Using (\ref{TCmu}), we defined the following truncated log-likelihood function under $\mathcal{H}_0$
\begin{equation}
\tilde{L}(\alpha,\beta,\theta)=\sum_{i=1}^{n}F\left(Y_i;\, \tilde{\mu}_{i}, \sigma\left(\tilde{\mu}_{i}\right)\right)
\label{TLL}
\end{equation}
It should be noted that (\ref{TLL}) is a log-likelihood function associated with a GMLM whose covariates are $Z_i$ and $C_i$, where $\widehat{\alpha}$, $\widehat{\beta}$ and $\widehat{\theta}$ are the maximum likelihood estimators (MLEs) of  $\alpha, \beta$ and $\theta$ respectively. Consequently, the MLE of the parameter function is given by:
\begin{equation*}
\widehat{\theta}(t)=\sum_{j=1}^{J}\widehat{\theta}_j\phi_j(t).
\end{equation*}
The quality of the estimation depends principally on $J$, i.e. the number of eigenfunctions used in the truncation strategy. This crucial parameter can be consistently chosen by inspecting the Akaike information criterion (AIC) related to (\ref{TLL}), as proved by \cite{muller2005generalized}. Note that we used a pre-selected $J$ based on the cumulative inertia. Indeed, we focused on the selection of a $J$ (using the AIC) with a cumulative inertia value below a given threshold (95\% in the present case)\cite{ahmed2018binary}.\\

\noindent  {\it\textbf{Estimation under the alternative hypothesis}}. Since the parameters $\beta$ and $\theta(\cdot)$ must be independent of the potential cluster, their estimates under $\mathcal{H}_0$ will be fixed in the alternative hypothesis $\mathcal{H}_1$. This means that under $\mathcal{H}_1$, covariate effects are invariant inside and outside the potential cluster. Hence, one only needs to estimate the parameters $\alpha$ and $\delta_k$ for each $S_k \in \mathcal{S}$. This can be achieved by maximizing the following log-likelihood function with respect to the two scalars:
\begin{equation}
\tilde{L}_k(\alpha,\delta_{k})=\sum_{i=1}^{n}F\left(Y_i;\, \tilde{\mu}_{i}^{(k)}, \sigma\left(\tilde{\mu}_{i}^{(k)}\right)\right),
\label{LLH1}
\end{equation} 
with 
$$\tilde{\mu}_{i}^{(k)}=\Phi^{-1}\left(\alpha+\delta_k\xi_{i}^{(k)}+Z_i^{'}\widehat{\beta}+\int_{\mathcal{T}} X_i(t)\widehat{\theta}(t)dt\right).$$
Let us consider $\widehat{\alpha}^{(k)}$ and $\widehat{\delta}_k$, denoting the MLEs of $\alpha_k$ and $\delta_k$, respectively. It should be noted that the covariate information is added as an offset, which illustrates the above-mentioned assumption concerning the independence of the potential cluster vs. the covariates. \\

\noindent {\it\textbf{Functional-model-adjusted spatial scan statistic.}} Using the MLEs determined under the two hypotheses, the LLR can be defined as follows:
\begin{equation}
\mathrm{LLR}_k=\tilde{L}_k\left(\widehat{\alpha}^{(k)},\widehat{\delta}_k\right)-\tilde{L}\left(\widehat{\alpha},\widehat{\beta},\widehat{\theta}\,\right).
\end{equation}
The MLC is then defined as the potential cluster $S_k$ that maximizes this ratio:
\begin{equation}
\mbox{MLC}=\mathrm{argmax}_{S_k \in \mathcal{S}}\{\mathrm{LLR}_k\}.
\end{equation}
Hence, the FMASSS is defined as the LLR associated with the MLC:
\begin{equation}
\lambda = \max_{S_k \in \mathcal{S}}\{\mathrm{LLR}_k\}.
\end{equation}
Since the distribution of $\lambda$ under $\mathcal{H}_0$ does not have a closed form, the significance of the MLC is evaluated by Monte-Carlo simulation. Each simulation $m$ $(m=1,\ldots,M)$ combines the real data (associated with the covariates) with a random dataset generated for the outcome variable. The latter is simulated using a conditional distribution under $\mathcal{H}_0$ (\textit{via} $\widehat{\alpha}, \widehat{\beta}$ and $\widehat{\theta}(\cdot)$). Let $\lambda^{(1)}, \dots ,\lambda^{(M)}$ denote the observations of the FMASSS on the simulated datasets. According to Dwass\cite{dwass1957modified}, the p-value of the FMASSS $\lambda$ observed in the real data is defined by $1-R/(M+1)$, where $R$ is the rank of $\lambda$ in the $(M+1)$-sample $\{\lambda^{(1)}, \dots ,\lambda^{(M)},\lambda\}$.\\

The FMASSS is built in three steps:\\

\noindent \underline{\textit{Construction of functional data, and dimension reduction}}
\begin{itemize}
	\item Construct the functional data by using a suitable basis of functions $\{\varphi_j(t),\, j=1,\ldots,K\}$. 
	\item Apply a functional PCA to the constructed functions $\{X_i(t), i=1,\ldots,n\}$. This is equivalent to a multiple PCA on the matrix $A\Psi^{1/2}$ where  $ \Psi$ is the $K\times K$ matrix with elements \citep{escabias2004principal} 
	$$\Psi_{jr}=\int_{\mathcal{T}}\varphi_j(t)\varphi_r(t)dt,\qquad j,r=1,\ldots,K.$$
	Thus, the eigenfunctions are defined by
	$\phi_{j}(t)=\sum_{j=1}^{K}v_{lj}\varphi_l(t), \quad j=1,\ldots,K$
	where $v_{lj}$ are the elements of the $K\times K$ matrix $V=\Psi^{-1/2}G$, where $G$is the eigenvector matrix associated with a multiple PCA of the matrix $A\Psi^{1/2}$. Moreover, the coefficients of the functional data in the eigenbasis are given by the $n\times K$ matrix  $C=A\Psi V$.
	\item Choose the optimal truncation parameter $J^{*}\in \{1,\ldots,K\}$,i.e. one that minimizes the AIC associated with models with the log-likelihood function defined in (\ref{TLL}).
\end{itemize}
\underline{\textit{Computation of the observed FMASSS}}
\begin{itemize}
	\item Use $J^{*}$ to estimate $\widehat{\alpha}$, $\widehat{\beta}$ and $\widehat{\theta}$ under $\mathcal{H}_0$ by using the log-likelihood function defined in (\ref{TLL}).
	\item For each potential cluster $S_k$, find $\widehat{\alpha}^{(k)}$ and $\widehat{\delta}_{k}$, that maximize (\ref{LLH1}) by adding the $Z_i\widehat{\beta}$ and $C_i^{'}\widehat{\theta}$ as offsets, then calculate the associated $\mathrm{LLR}_k$. Moreover, identify the MLC and its FMASSS $\lambda$, over the set of potential clusters.
\end{itemize}	
\underline{\textit{Monte-Carlo simulation}}
\begin{itemize}
	\item Apply the Monte-Carlo hypothesis testing procedure described above.
\end{itemize}

\section{Application to a Poisson model}
\label{Poisson}
This section describes the estimation procedure when the data on the outcome variable $Y$ have a Poisson distribution. Let $N_i$ be the measurement of the underlying at-risk population associated with the $i$th location $s_i$. The Poisson model is characterized by the following link function $\Phi(\cdot)$ and the conditional log-likelihood (\ref{LL}):
\begin{equation}
\Phi(t)=\log(t) \qquad \mbox{and}\quad F\left(Y_i;\, \mu_{i}^{(k)}\right)=Y_i\log\left(\mu_{i}^{(k)}\right)-\mu_{i}^{(k)}-\log\left(Y_i!\right),
\label{poissonLL}
\end{equation}
with 
\begin{equation*}
\mu_i^{(k)}=E\left(Y_i\left\vert S_k,Z_i,X_i\right.\right)=N_i\exp\left(\alpha+\delta_{k}\xi_{i}^{(k)}+Z_i^{'}\beta+\int_{\mathcal{T}} X_i(t)\theta(t)dt\right).
\end{equation*}
It should be noted that multiplication by $N_i$ makes it possible to take account of the underlying at-risk population as an adjustment covariate. Consequently, $\log(N_i)$ is taken as an offset in the model. Let $\widehat{\alpha}$, $\widehat{\beta}$ and $\widehat{\theta}(\cdot)$ be the MLEs under the null hypothesis. It can be shown that the MLE $\widehat{\alpha}$ is expressed in the following manner (for details, see the Appendix):
$$\widehat{\alpha}=\log\left(\frac{\sum_{i=1}^{n}Y_i}{\sum_{i=1}^{n}\tilde{N}_i}\right),\quad \mbox{where}\quad   \tilde{N}_i=N_i\exp\left(Z_i^{'}\widehat{\beta}+\int_{\mathcal{T}} X_i(t)\widehat{\theta}(t)dt\right).$$
Note that the $\exp(\widehat{\alpha})$ can be viewed as the incidence rate under $\mathcal{H}_0$ in the adjusted underlying at-risk population $\tilde{N}_i$ rather than in the initial underlying at-risk population $N_i$.\\
As detailed in section \ref{Methodology}, the estimation procedure under $\mathcal{H}_1$ consists in maximizing the log-likelihood (\ref{LLH1}), which is expressed as follows:
\begin{equation*}
\tilde{L}_k(\alpha,\delta_{k})=\sum_{i=1}^{n}Y_i\left(\alpha+\delta_k\xi_{i}^{(k)}+\log\left(\tilde{N_i}\right)\right)-\tilde{N_i}\exp\left(\alpha+\delta_k\xi_{i}^{(k)}\right)-\log\left(Y_i!\right),
\end{equation*} 
taking its maximum at:
\begin{equation*}
\widehat{\alpha}^{(k)}=\log\left(\frac{\sum_{i=1}^{n}Y_i(1-\xi_{i}^{(k)})}{\sum_{i=1}^{n}\tilde{N}_i(1-\xi_{i}^{(k)})}\right) \qquad \mbox{and} \qquad \widehat{\delta}_k=\log\left(\frac{\sum_{i=1}^{n}Y_i\xi_{i}^{(k)}}{\sum_{i=1}^{n}\tilde{N}_i\xi_{i}^{(k)}}\frac{\sum_{i=1}^{n}\tilde{N}_i(1-\xi_{i}^{(k)})}{\sum_{i=1}^{n}Y_i(1-\xi_{i}^{(k)})}\right).
\end{equation*}
It should be noted that $\exp(\widehat{\delta}_k)$ is the relative risk associated with the potential cluster $S_k$ after adjusting for the underlying at-risk population $\tilde{N}_i$.\\

Next, $\mathrm{LLR}_k$ is given by:
\begin{eqnarray}
\mathrm{LLR}_k&=&\tilde{L}_k(\widehat{\alpha}^{(k)},\widehat{\delta}_k)-\tilde{L}(\widehat{\alpha},\widehat{\beta},\widehat{\theta}\,) \nonumber\\
&=&\left[O^{(k)}\log\left(\frac{O^{(k)}}{\tilde{N}^{(k)}}\right)+(O-O^{(k)})\log\left(\frac{O-O^{(k)}}{\tilde{N}-\tilde{N}^{(k)}}\right)\right]-O\log\left(\frac{O}{\tilde{N}}\right),
\label{LLRPois}
\end{eqnarray}
where 
$$\tilde{N}=\sum_{i=1}^{n}\tilde{N}_i\,, \qquad \tilde{N}^{(k)}=\sum_{i=1}^{n}\tilde{N}_i\xi_{i}^{(k)}\,,\qquad O=\sum_{i=1}^{n}Y_i \qquad \mbox{and}\qquad O^{(k)}=\sum_{i=1}^{n}Y_i\xi_{i}^{(k)}.$$
It should be noted that ($\ref{LLRPois}$) is equivalent to the LLR proposed by \cite{kulldorff1997spatial} for a Poisson model, except that the adjusted underlying at-risk population $\tilde{N}_i$ is taken into account (rather than $N_i$). In other words, adjustment for covariates is equivalent to considering a Poisson model with an underlying at-risk population adjusted under the null hypothesis.

\section{Simulation study}
\label{Simulation_study}
We simulated a cluster detection procedure in order to compare the quality of adjustment for a longitudinal confounding factor in three spatial scan statistic models: a univariate model, a multivariate model, and a functional model.
\subsection{Design of the simulation}
Artificial datasets were generated according to Poisson models by using the geographic locations of the $n = 94$ French administrative areas (\textit{d\'epartements}, as shown in Figure~\ref{FigCarte} in the Supplementary Material) and population data from the French national census database (\textit{Institut National de la Statistique et des Etudes Economiques}, INSEE). Each location was defined as the \textit{d\'epartement}'s administrative center. Two types of non-overlapping cluster (each containing 8 \textit{d\'epartements}) were defined and simulated for each artificial dataset. The first was entirely characterized by the cluster intensity $\delta$, namely the \textit{true cluster} (the areas in green in Figure~\ref{FigCarte}), and the second was characterized solely by the effect of the functional covariate, namely the \textit{fake cluster} (the areas in red in Figure~\ref{FigCarte}).\\

\noindent{\it \textbf{Generation of the artificial datasets}}. The random functions were simulated as the realization of the following process in the interval $[0, 21]$:   
\begin{equation*}
X_i(t)=\left\{
\begin{array}{lcl}
U_ih(t)+\left(1-U_i\right)h(t+4) + \epsilon_{i}(t) & & \mbox{for $s_i$  outside the \textit{fake cluster}} \\ 
& & \\ 
U_ih(t)+\left(1-U_i\right)h(t-4) +\epsilon_i(t) & & \mbox{for $s_i$ inside the \textit{fake cluster}} \\ 
\end{array}
\right.
\end{equation*}
where $h(t)=\max\left(6-|t-11|,0\right)$, $U_i$ is uniform, and $\epsilon_i(t)$ are uncorrelated, normally distributed random variables. A total of 94 curves were simulated with respect to the random function $X(\cdot)$ (Figure~\ref{Curves}, left panel) and used to generate data from the following Poisson model:

\begin{figure}[!h] 
	\centering
	\includegraphics[width=1\textwidth]{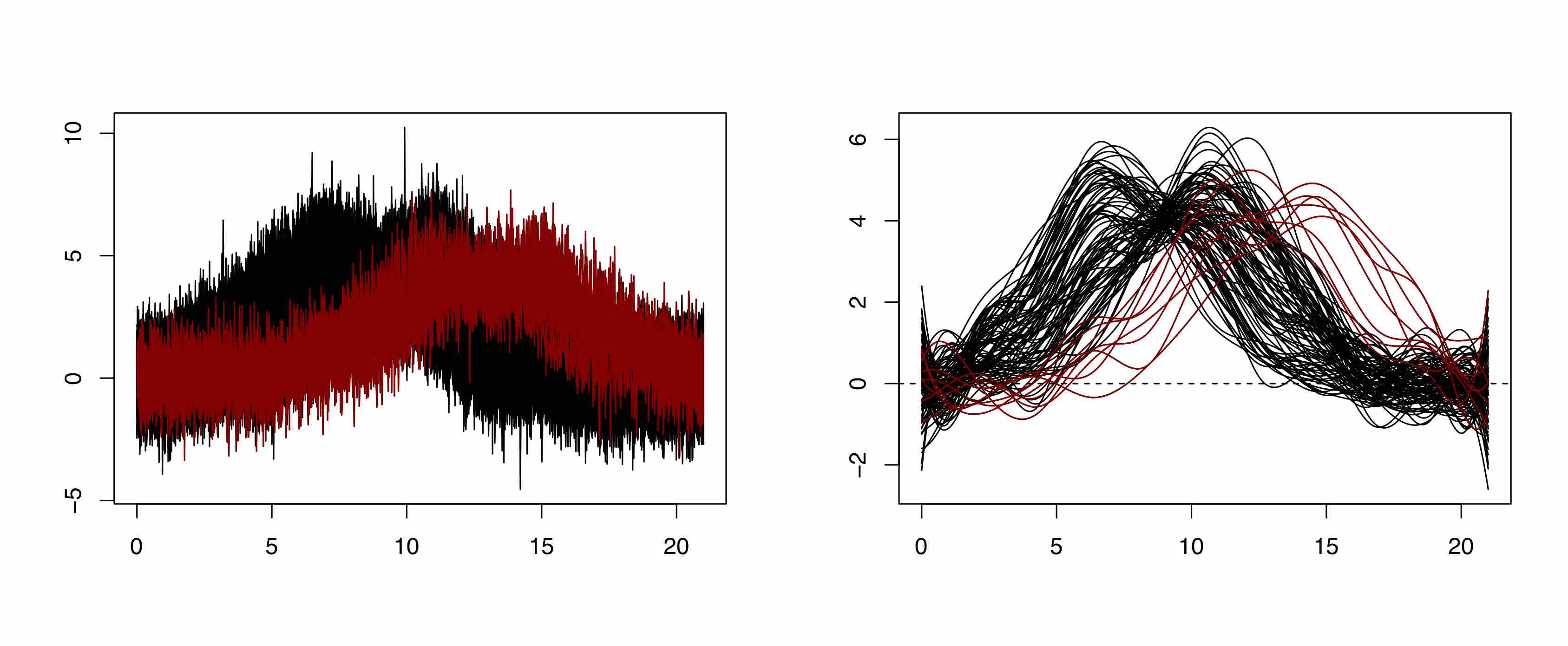}
	\caption{Simulation study: an example of the generated longitudinal data before (left panel) and after (right panel) smoothing. The red curves correspond to the observations in the \textit{fake cluster}.}
	\label{Curves}
\end{figure}
\begin{equation}
\mu_i=N_i\exp\left(\alpha+\delta\xi_{i}+\int_{0}^{21} X_i(t)\theta(t)dt\right)
\label{Poi}
\end{equation}
where $N_i$ corresponds to the at-risk population in the $i$th \textit{d\'epartement} and $\xi_i=1$ for \textit{d\'epartements} located in the \textit{true cluster}.
Firstly, an intercept $\alpha=-11.51$ was chosen to ensure a disease incidence of approximately $10^{-5}$ in the absence of a cluster and the absence of a confounding covariate. Secondly, the confounding functional covariate was introduced into the model using $\theta(t)=\frac{t}{9}\sin(\frac{\pi}{9}t+\pi),\, t\in [0, 21]$ in such a way that the mean value of the outcome was twice as high inside the \textit{fake cluster} as outside. Thirdly, different values of the \textit{true cluster} intensity were considered and expressed in terms of the relative risk: $\exp(\delta) \in \{1, 1.2, 1.4, 1.6, 1.8, 2\}$. \\

\noindent{\it \textbf{Comparison of three models}}. To illustrate the performance of the functional approach to adjustment, we compared three models. We considered that for each location, the functional covariate was only observed, at 70 time points equally spaced throughout the interval $[0, 21]$. Below, the term "longitudinal data" refers to the realization of the functional covariate at these 70 time points.

In the univariate model, the outcome variable was adjusted by a single covariate (the average of the longitudinal data). In the multivariate model, the outcome variable was adjusted by 70 random covariates with the values of the 70 time points by using Jung's method\cite{jung2009generalized}. In order to deal with the strong collinearity between these 70 covariates, a multiple PCA was applied by using the AIC-based selection method described in Section  \ref{Methodology}. Lastly, in the functional model, the outcome variable was adjusted by using the smoothed curves as a functional covariate. The latter was constructed from the longitudinal data by using a cubic B-spline basis of functions, as defined by 13 equally spaced knots in the interval $[0, 21]$ (the right panel in Figure~\ref{Curves}).

For each value of the cluster intensity, $1000$ artificial datasets were simulated. The three models were compared with regard to three distinct criteria: the power to detect a significant cluster (\textit{true} or \textit{fake}), the true-positive (TP) rate, and the false-positive (FP) rate. The power of each model was defined as the proportion of datasets highlighting a significant cluster (a \textit{true} or \textit{fake} cluster), with a type I error of $0.05$ and $999$ Monte-Carlo simulations. The TP and FP rates were calculated according to Cucala et al.'s method\cite{cucala2018multivariate}.

\subsection{Results of the simulation study}
The results of the simulation study are shown in Figure~\ref{TC_FC_simul}  (see Table~\ref{RSim} in the supplementary material for more details). The adjustment based on a univariate model (with the average of the longitudinal data as a covariate) failed to detect the \textit{true cluster} as the MLC. This was particularly the case for cluster intensity values that were low or moderate, relative to the intensity of the \textit{fake cluster}. The univariate model detected the \textit{fake cluster} as the MLC, as illustrated by the curves for the power and the TP and FP rates in Figure~\ref{TC_FC_simul}. The adjustments based on the functional and multivariate models did not differ significantly with regard to the power or the TP and FP rates for detecting the \textit{true cluster}. The functional model performed slightly better for high cluster intensities ($\exp(\delta)=1.8 \mbox{ and } 2.0$). As expected, the power of both models increased with the cluster intensity. It can be seen that both the multivariate model and the functional model seldom detected the \textit{fake cluster} as the MLC.
\begin{figure}
	\centering
	\includegraphics[width=0.9\textwidth]{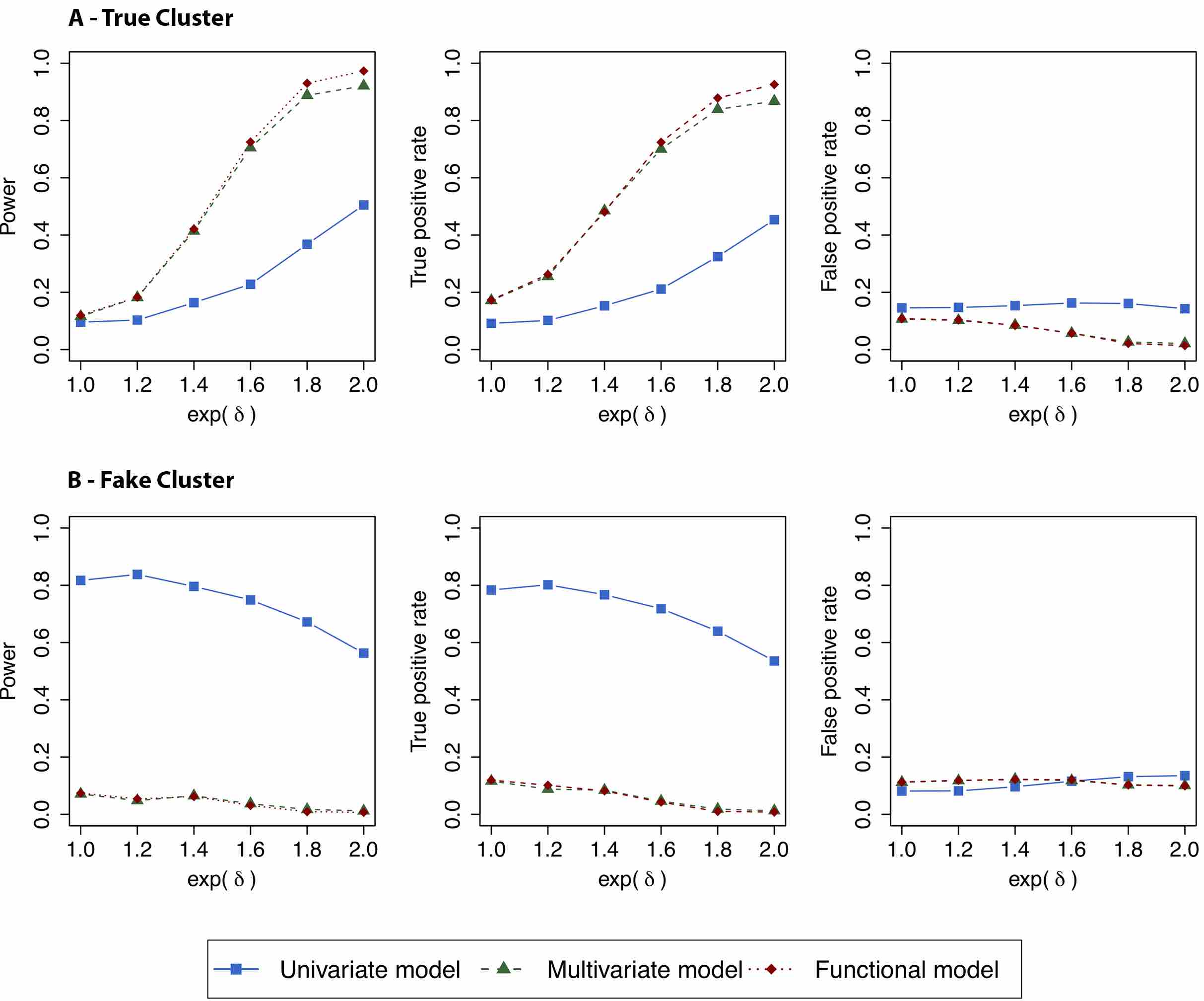}
	\caption{Simulation study: comparison of the univariate, multivariate and functional models with regard to the quality of adjustment. For each model, the power curves and the true-positive and false-positive rates for the detection of the \textit{true cluster} (A) and the \textit{fake cluster} (B) as most likely cluster are shown. The quantity $\exp{(\delta)}$ refers to the cluster intensity.}
	\label{TC_FC_simul}
\end{figure}

\section{Application to epidemiologic data}
\label{Real_data}
\subsection{Premature mortality and related confounding factors}
We considered data provided by the INSEE on premature mortality in France between 1998 and 2013. Premature mortality was defined as death before the age of 65. For each of the 94 French \textit{d\'epartements} (administrative areas) and for the period between 1998 and 2013, the mean premature mortality rate was defined as the number of persons who died before the age of 65, divided by the mean number of persons aged under 65. Hereafter, the outcome variable refers to the number of premature deaths per \textit{d\'epartement} between 1998 and 2013. The spatial distribution of premature mortality in France is shown in Figure \ref{Premature_graph} (supplementary materials).\

\begin{figure}[!h] 
	\centering
	\includegraphics[width=0.49\textwidth]{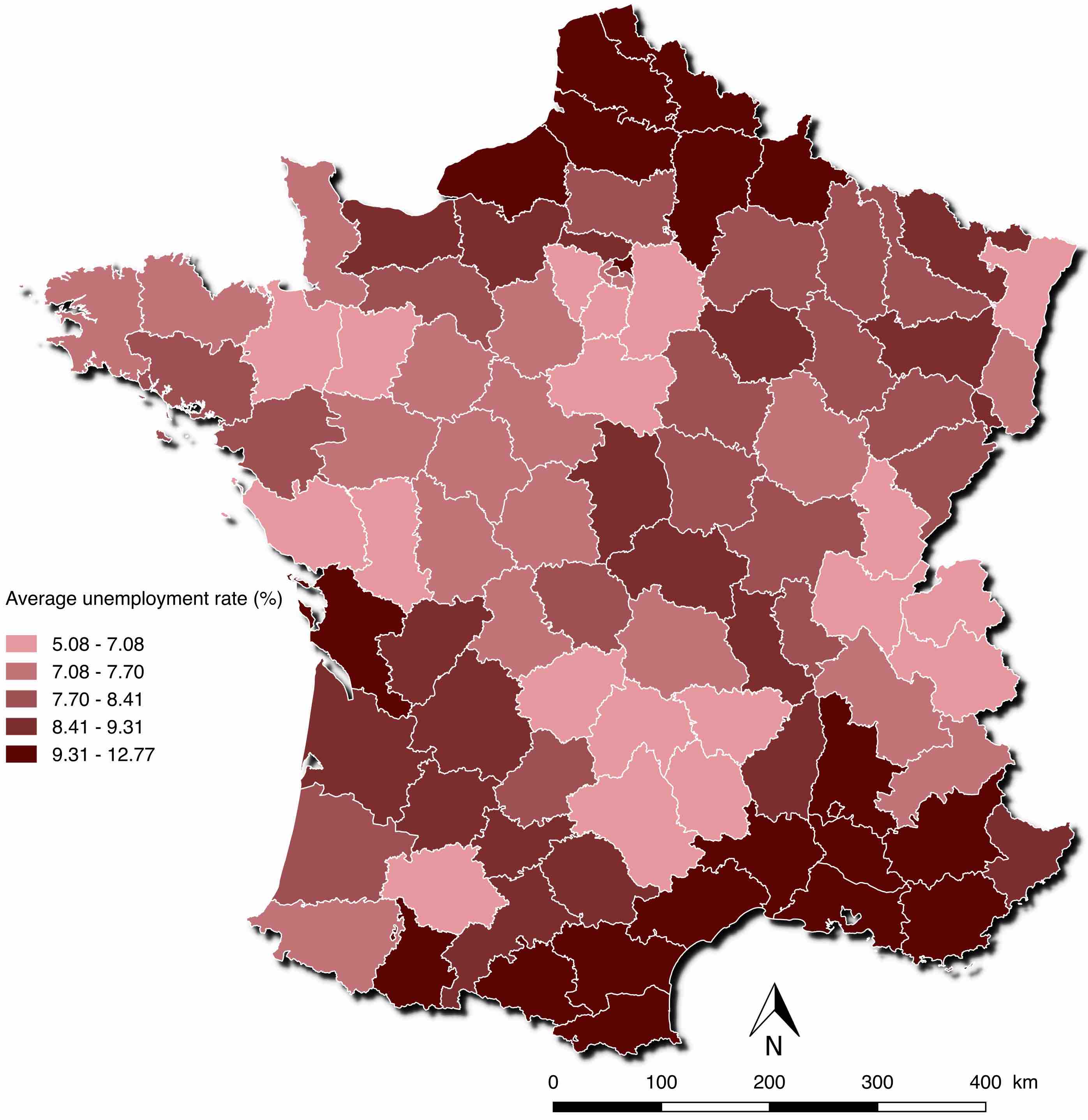}
	\includegraphics[width=0.49\textwidth,height=0.5\linewidth]{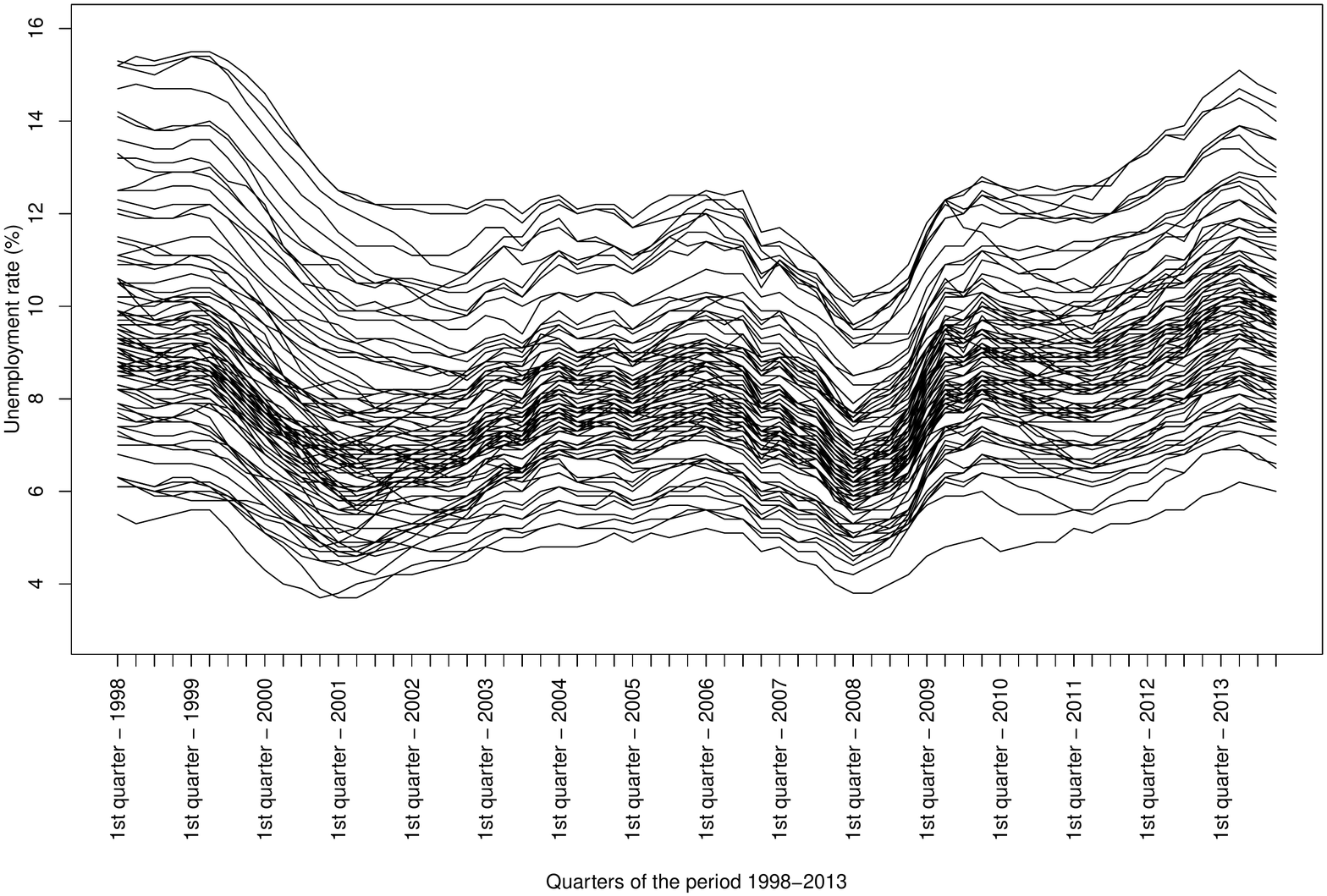} 
	\caption{The unemployment rate in France from 1998 to 2013, by \emph{d\'epartement}. The left panel shows the unemployment rate averaged over the 16-year period from 1998 to 2013 for each \emph{d\'epartement}. The right panel shows the change over time in the unemployment rate between 1998 and 2013; each curve corresponds to a \emph{d\'epartement}.}
	\label{Unemployment rate}
\end{figure}

It is known that premature mortality affects men more than women, and is correlated with socio-economic status: the most deprived are more likely to die young\cite{STRINGHINI20171229}. Thus, it is important to adjust the spatial cluster detection analyzes for the confounding factors of gender and socio-economic status. To this end, we considered the mean proportion of men aged under 65 over the period from 1998 to 2013 for each \textit{d\'epartement} (as provided by the INSEE database). We chose the mean proportion because it did not greatly vary over the 16-year period (see  Figure \ref{Sex_graph} in the supplementary materials). We considered the unemployment rate (in \%) for each quarter of the period from 1998 to 2013 as a proxy for socioeconomic status - leading to 64 values per \textit{d\'epartement}. Figure \ref{Unemployment rate} shows both the spatial distribution of the mean unemployment rate over the entire period and the change over time in the unemployment rate for each of the \textit{d\'epartements}. The mean unemployment rate is spatially heterogeneous. Furthermore, the unemployment rate varied markedly between 1998 and 2013, and thus must be considered as a longitudinal confounding factor.

\subsection{Spatial clusters detection}
In order to detect spatial clusters of premature mortality, four different Poisson models were considered. Each model was adjusted for gender by introducing the mean proportion of men by \emph{d\'epartement} over the period from 1998 to 2013 as a covariate. The four models are described below:
\begin{enumerate}
\item Model 1 (the non-adjusted model): no adjustment of the outcome variable for the unemployment rate.
\item Model 2 (the univariate model): adjustment of the outcome variable for the unemployment rate, using the mean rate over the period from 1998 to 2013 by \emph{d\'epartement} as a single covariate.
\item Model 3 (the multivariate model): adjustment of the variable outcome for the unemployment rate by considering the each of the quarterly values by \emph{d\'epartement} for the period from 1998 to 2013 as a covariate. Thus, 64 covariates related to the unemployment rate were introduced into the model.
\item Model 4 (the functional model): adjustment of the outcome variable for the unemployment rate using smoothed rate curves as a functional covariate. The curves were built from the data using a cubic B-spline basis defined by 15 knots in the interval $[0, 1]$. The least-squares method was used to compute the corresponding coefficients for each random curve.
\end{enumerate}
Each model was used to detect spatial clusters with a high-risk of premature mortality (i.e. with a relative risk (RR) $=\exp(\delta)>1$) or with a low risk of premature mortality (RR $=\exp(\delta)<1$). The MLC was considered, together with secondary clusters that had a high FMASSS value and did not cover the MLC \cite{kulldorff1997spatial}. The statistically significance of the detected spatial clusters was evaluated by performing 999 Monte-Carlo simulations, with a type I error of 0.05.

\subsection{Results}

\begin{figure}
\centering
\includegraphics[width=0.9\textwidth]{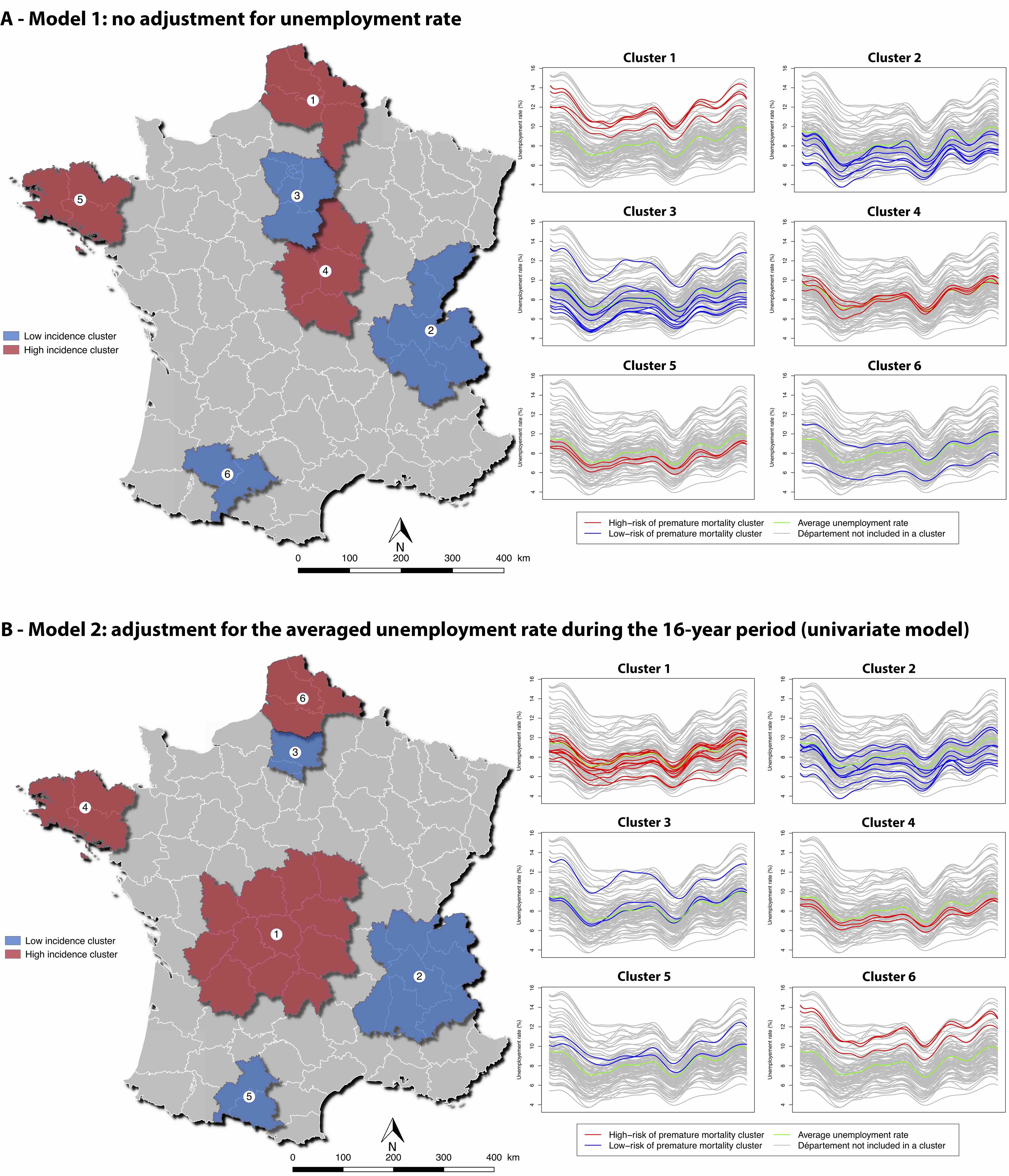}\\
\caption{Significant spatial clusters of premature mortality detected by the model not adjusted for the unemployment rate (top panel A) and the univariate model (bottom panel B). Spatial clusters in red indicate a high risk of premature mortality, and those in blue show indicate a low risk of premature mortality. The clusters are numbered as follows: cluster 1 is the most likely cluster, and the other (secondary) clusters a number in descending order for their test statistic. For each cluster, the unemployment rate curves (from 1998 to 2013) in each \emph{d\'epartement} are presented. Curves in blue correspond to \emph{d\'epartements} inside the cluster, curves in red correspond to \emph{d\'epartements} outside the cluster, and the curve in green is the mean curve.}
\label{maps_clusters_noadjust_univariate}
\end{figure}

\begin{figure}
\centering
\includegraphics[width=0.9\textwidth]{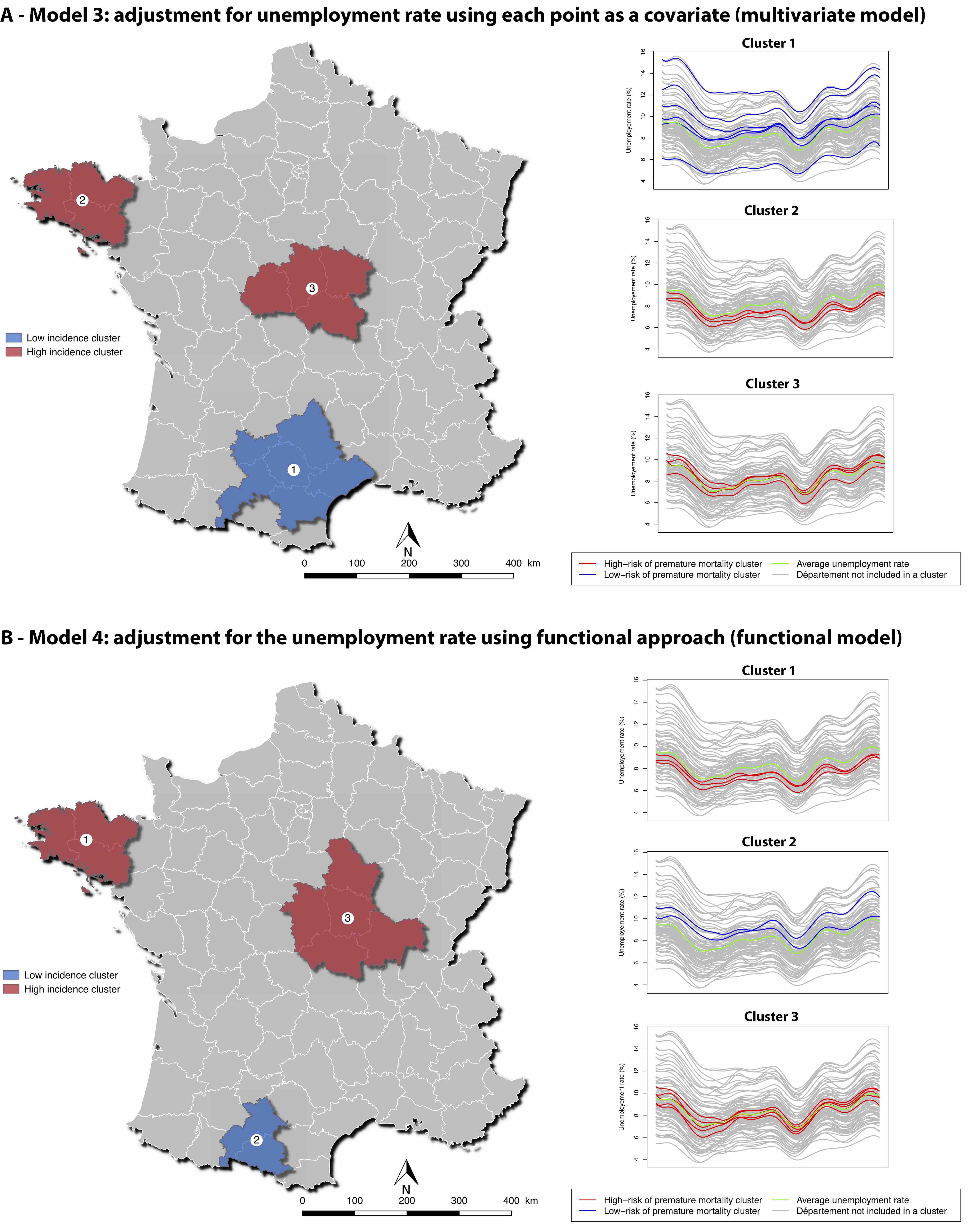}\\
\caption{Significant spatial clusters of premature mortality detected by the multivariate model (top panel A) and the functional model (bottom panel B). Spatial clusters in red color indicate a high risk of premature mortality, and those in blue indicate a low risk of premature mortality. The clusters are numbered as follows: cluster 1 is the most likely cluster, and the other (secondary) clusters a number in descending order for their test statistic. For each cluster, the unemployment rate curves (from 1998 to 2013) in each \emph{d\'epartement} are presented. Curves in blue correspond to \emph{d\'epartements} inside the cluster, curves in red correspond to \emph{d\'epartements} outside the cluster, and the curve in green is the mean curve.}
\label{maps_clusters_multivariate_functional}
\end{figure}

The statistically significant spatial clusters detected by the non-adjusted and univariate models (models 1 and 2) are presented in Figure \ref{maps_clusters_noadjust_univariate}, and those identified by the multivariate and the functional models (models 3 and 4) are displayed in Figure \ref{maps_clusters_multivariate_functional}. Detailed information on the spatial clusters is presented in Table \ref{Tab_cluster_premature_mortality}.

\begin{table}
	\centering
\caption{Statistically significant spatial clusters of premature mortality detected in the absence of adjustment for the unemployment rate (Model 1), a univariate model (Model 2), a multivariate model (Model 3) and a functional model (Model 4).}
\label{Tab_cluster_premature_mortality}
\begin{tabular}{l c c c  c c}
\toprule
Model&  Cluster & \# \emph{d\'epartements}  & relative risk $(\exp(\delta))$ & $\mathrm{LLR}$ & P-value  \\
\midrule 
\multirow{5}{*}{Model 1}
& 1  &  4 &1.28 &4648.24 &0.001 \\
&2  &  7 & 0.79& 3225.33& 0.001 \\
&3  &  9 & 0.86 &2939.85 &0.001 \\
&4  &  4 &1.28 &1131.82 &0.001 \\
&5  &  3 &1.18 & 856.63& 0.001 \\
&6  &  2 &0.80 & 827.15 &0.001\\
\midrule
\multirow{3}{*}{Model 2}
&1 &  12 &1.19 &1531.91& 0.001\\
&2  &  8 &0.86& 1458.09& 0.001 \\
&3  &  3 &0.85 &1120.88 &0.001 \\
&4  &  3 &1.20 &1091.54 &0.001 \\
&5  &  2 &0.77 &1090.26 &0.001 \\
&6  &  3& 1.08 & 405.53 &0.001\\
\midrule
\multirow{3}{*}{Model 3}
&1  &  6 &0.86 &916.17 &0.001\\
&2 &   3 &1.17 &795.61 &0.001\\
&3  &  4 & 1.19& 511.30& 0.001\\
\midrule
\multirow{3}{*}{Model 4}
&1 &   3 & 1.24 &1455.90& 0.001\\
&2 &   2& 0.74& 1398.57& 0.001\\
&3  &  5&  1.21&917.15 & 0.001 \\ 
\bottomrule
\end{tabular}
\end{table}

Model 1 identified 6 significant spatial clusters of premature mortality: 3 low-risk clusters (RR: 0.79 to 0.86) and 3 high-risk clusters (RR: 1.18 to 1.28) (top panel in Figure \ref{maps_clusters_noadjust_univariate}). The MLC (Cluster 1, RR=1.28) was located in northern France, and was characterized by a high unemployment rate. Similarly, the first secondary cluster (Cluster 2, RR=0.79) was located in eastern France and was characterized by a low unemployment rate. 

Model 2 also identified 6 significant spatial clusters of premature mortality: 3 low-risk clusters (RR: 0.77 to 0.86) and 3 high-risk clusters (RR: 1.08 to 1.20) (bottom panel in Figure  \ref{maps_clusters_noadjust_univariate}). Like model 1, model 2 also detected the cluster with a high unemployment rate in northern France (Cluster 6, RR: 1.08) and the cluster with a low unemployment rate in eastern France (Cluster 2, RR=0.86) - emphasizing the poor quality of adjustment when using solely the mean unemployment rate over the study period. 

Model 3 detected 3 statistically significant spatial clusters of premature mortality: a low-risk cluster (RR: 0.86) and 2 high-risk clusters (RR: 1.17 and 1.19, respectively) (top panel in Figure \ref{maps_clusters_multivariate_functional}). It should be noted that the clusters characterized by a high or low unemployment rate (in northern and eastern France, respectively) detected by models 1 and 2 were not detected by model 3. The MLC in model 3 (Cluster 1; RR: 0.86) highlighted significant heterogeneity in the unemployment rates because it included a \textit{d\'epartement} with a high unemployment rate and a \textit{d\'epartement} with a low unemployment rate. 

Model 4 highlighted 3 statistically significant spatial clusters of premature mortality: 1 low-risk cluster (RR: 0.74) and 2 high-risk clusters (RR: 1.21 and 1.24, respectively) (bottom panel in Figure \ref{maps_clusters_multivariate_functional}). These three clusters are characterized by unemployment rate curves close to the average curve (in green). This result shows that the cluster detection was well adjusted for the unemployment rate.

\section{Discussion}
\label{Discussion}
Here, we developed an FMASSS in order to adjust cluster detection for longitudinal confounding factors in a purely spatial analysis. In other words, we addressed the issue of adjusting a spatial scan statistic for repeatedly measured covariates whose values vary over time. The FMASSS was derived by modeling the longitudinal confounding factor as a random function. The corresponding basis of functions depends principally on the nature of the longitudinal data. One advantage of using a random function is its consideration of the entire set of longitudinal data, rather than a rough approximation by a statistical indicator such as the mean (which is often the case in spatial epidemiological studies). Furthermore, this functional approach makes it possible to overcome (i) the missing data problem related to the difference in measurement times between spatial units, and (ii) the high dimensionality inherently associated with multivariate approaches when longitudinal data are measured at many time points. Our approach was built into a general framework for use with various parametric models (Bernoulli, Gaussian, and Poisson models, etc.). For a Poisson model, it has been shown that the FMASSS is equivalent to Kulldorff's classical spatial scan statistic in an adjusted population \citep{kulldorff1997spatial}.\\

We next simulated and compared different way of adjusting the spatial scan statistics for longitudinal confounders. The univariate model did not adjust the data well, and detected a \textit{fake cluster} when the cluster intensity was weak or moderate. In contrast, the multivariate and functional models were both able detect the \textit{true cluster} with a high power. The functional model was slightly better than the multivariate model for high cluster intensities. It should be noted that this general power equivalence for the two latter models is partly due to the design of the simulation study. In fact, the simulation represented an ideal situation because the measurement times for the longitudinal data were the same in all the spatial units; hence, there were no missing data in the multivariate model.\\

These models were applied to the detection of spatial clusters of premature mortality in France over the period from 1998 to 2013. The proportion of men by \textit{d\'epartement} and the unemployment rates for each quarter of the study period (64 values per \textit{d\'epartement}) were considered as confounding variables. The clusters considered to be significant in the univariate model (based on the mean unemployment rate over the entire study period) were characterized by unemployment rates that were far from the mean. This finding highlighted the univariate model's poor ability to adjust for a longitudinal confounding factor summarized as the mean. In the multivariate model, the MLC also included \textit{d\'epartements} with unemployment rates that were far from the mean value - again showing that the adjustment was not optimal. In contrast, the spatial clusters of premature mortality detected by the functional model had unemployment rates that were very close to the mean - testifying to high-quality adjustment for the longitudinal confounding factor. In the present application, it would have been interesting to adjust to environmental factors that are usually measured daily or weekly. The new method presented here is very well suited to this type of longitudinal data.\\

It should be borne in mind that the new FMASSS deals with round-shaped clusters only (the simplest case). However, clusters may be elongated in some situations - such as the aggregation of cases of water-borne disease along a river. However, the FMASSS can easily be extended to other spatial cluster shapes, such as elliptical clusters \citep{kulldorff2006elliptic}, graph-based clusters \citep{cucala2013spatial} or (for a spatiotemporal framework) cylindrical clusters \citep{kulldorff2005space}. \\

Lastly, the FMASSS can be extended to spatiotemporal frameworks in which the outcome measure and longitudinal confounders are measured on different time scales (e.g. an outcome measured annually and a longitudinal confounding factor measured monthly). In this context, the longitudinal data can be represented by a random function for each of the outcome time units.

%
%
%
%

\bibliographystyle{model5-names}
\bibliography{biblio}%

\appendix
\section{An explicit intercept estimator in the Poisson model under $\mathcal{H}_0$}
\label{Appendix}
Under the null hypothesis, the truncated log-likelihood function (\ref{TLL}) associated with the Poisson model (\ref{poissonLL}) is given by:
\begin{eqnarray*}
\tilde{L}(\alpha,\beta,\theta)&&\sum_{i=1}^{n}Y_i\left(\log(N_i)+\alpha+Z_i^{'}\beta+ C_i^{'}\theta\right)\\
&&\qquad\qquad\qquad-\sum_{i=1}^{n}N_i\exp\left(\alpha+Z_i^{'}\beta+ C_i^{'}\theta\right)-\sum_{i=1}^{n}\log(Y_i!).
\end{eqnarray*}
and has the following first partially derivative with respect to $\alpha$: 
\begin{equation*}
\frac{\partial \tilde{L}}{\partial \alpha}(\alpha,\beta,\theta)=\sum_{i=1}^{n}Y_i -\sum_{i=1}^{n}N_i\exp\left(\alpha+Z_i^{'}\beta+ C_i^{'}\theta\right). 
\end{equation*}
It should be borne in mind that the MLEs $\widehat{\alpha}$, $\widehat{\beta}$ and $\widehat{\theta}$ of $\alpha$, $\beta$ and $\theta$, respectively, have to satisfy the first-order condition:
 \begin{equation*}
\frac{\partial\tilde{L}}{\partial \alpha}\left(\widehat{\alpha},\widehat{\beta},\widehat{\theta}\right)=\sum_{i=1}^{n}Y_i -\sum_{i=1}^{n}N_i\exp\left(\widehat{\alpha}+Z_i^{'}\widehat{\beta}+ C_i^{'}\widehat{\theta}\right)=0. 
 \end{equation*}
 Therefore, the coefficient $\widehat{\alpha}$ has an explicit expression with respect to the other coefficients $\widehat{\beta}$ and $\widehat{\theta}$
 \begin{equation*}
 \exp\left(\widehat{\alpha}\right)=\frac{\sum_{i=1}^{n}Y_i}{\sum_{i=1}^{n}\tilde{N}_i}, \qquad \mbox{with}\qquad \tilde{N}_i=N_i\exp\left(Z_i^{'}\widehat{\beta}+ C_i^{'}\widehat{\theta}\right).
 \end{equation*}
\begin{flushright}
$\square$
\end{flushright}

\section{Supplementary material}
\begin{figure}
	\centering
	\includegraphics[width=0.5\textwidth]{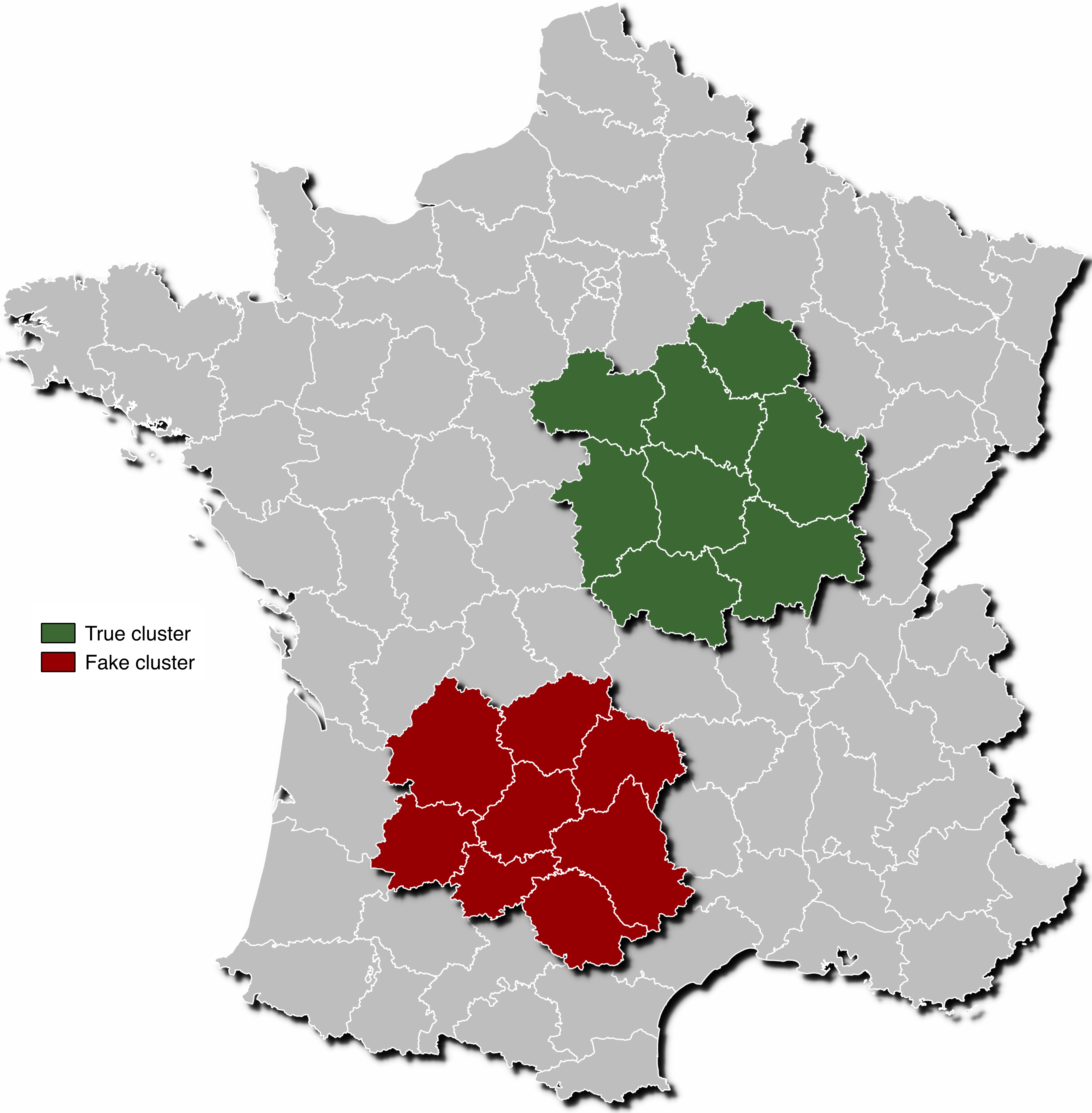}
	\caption{Simulation study: the \textit{true} and \textit{fake} simulated clusters among the 94 \emph{d\'epartements} (administrative areas) of France.}
	\label{FigCarte}
\end{figure}

\begin{table}
\caption{Simulation study: comparison of the quality of adjustment using functional, multivariate, and univariate models. For each model, the power, the true-positive rate and the false-positive rate related to the detection of the MLC as a \textit{true cluster} or a \textit{fake cluster} are given.}
	\label{RSim}
	\begin{center}
		\begin{tabular}{l l ccc cccc  cccc}
			\toprule
			\multirow{2}{*}{$\exp(\delta)$} & 	\multirow{2}{*}{Cluster}  & \multicolumn{4}{c}{Univariate model}& \multicolumn{4}{c}{Multivariate model}  & \multicolumn{3}{c}{Functional model} \\ \cline{3-5} \cline{7-9} \cline{11-13}
			&          & Power & TP   & FP   &   &  Power & TP  & FP   &  & Power & TP  & FP \\
			\midrule
			1         & True     &0.012  &0.014 & 0.100 &  & 0.075 &0.169& 0.104& & 0.083 &0.183& 0.109\\
			          & Fake     &0.914 &0.853& 0.022 &    &{\bf0.039} &{\bf0.10}6 &{\bf0.110}&  & 0.041 &0.108 &0.116 \\
			          \midrule
			1.2       & True     &0.034 &0.037& 0.103 &    &{\bf0.143}& 0.274 &0.100 & &0.142 &{\bf0.277} &{\bf0.099} \\
			          &  Fake    &0.893 &0.857& 0.026 &    &0.037 &0.090 &0.117&  & 0.041 &0.097 &0.116 \\
			          \midrule
			1.4       & True     & 0.093& 0.084& 0.105&    & 0.344 &{\bf0.494}& 0.084& & {\bf0.347} &0.488 &{\bf0.081} \\
			          & Fake     &0.862 &0.813& 0.037 &    &0.044& 0.074 &0.123&  &0.039 &0.069 &0.120 \\
			          \midrule
			1.6      &  True     & 0.248 &0.224 &0.102&& 0.656& 0.709& {\bf0.052}&& {\bf0.662}& {\bf0.712} &0.055 \\
			         & Fake      &0.760 &0.725 &0.055 &&0.034 &0.043& 0.114&& 0.036& 0.046 &0.117 \\ 
			         \midrule
			1.8      & True      &0.453 &0.424 &0.084&& 0.869& 0.841& 0.029&& {\bf0.890} &{\bf0.853}& {\bf0.026} \\
			         & Fake      &0.588 &0.553 &0.072&& 0.021& 0.023& 0.105&& 0.015& 0.018& 0.104 \\ 
			         \midrule
 			2.0     & True      &0.676 &0.642 &0.054 &&0.948 &0.898& 0.018& & {\bf 0.977} & {\bf0.929} & {\bf0.013} \\
			         & Fake      &0.366 &0.345& 0.081&& 0.008& 0.011& 0.100 &&0.006& 0.007 &0.099\\
			         \bottomrule
				\end{tabular}	
	\end{center}
\end{table}

\begin{figure}
	\centering
	\includegraphics[width=0.5\textwidth]{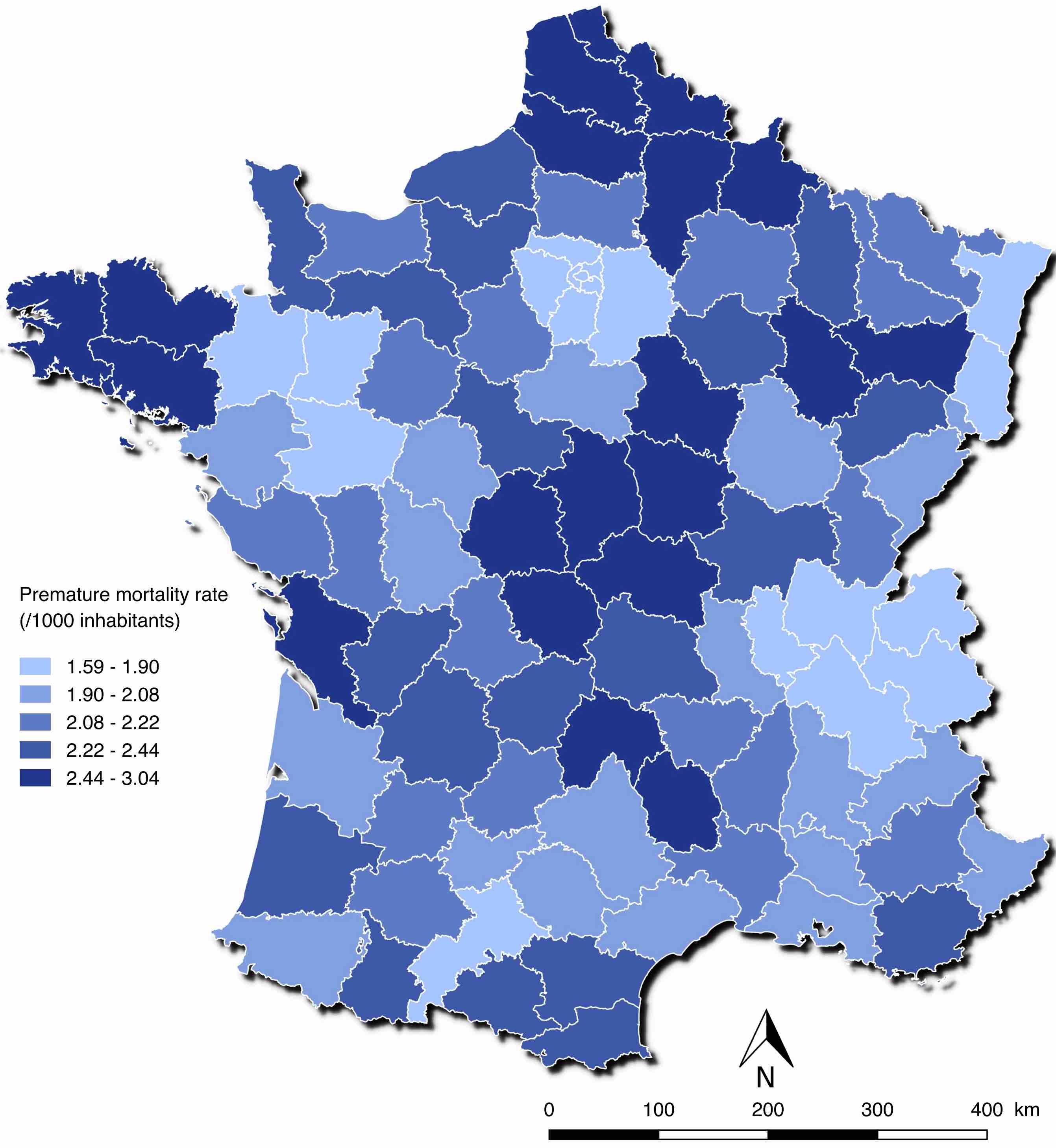} 
	\caption{Spatial distribution of the cumulative incidence of premature mortality in France during the period from 1998 to 2013.}
	\label{Premature_graph}
\end{figure}
\begin{figure}
	\centering
	\includegraphics[width=0.7\textwidth]{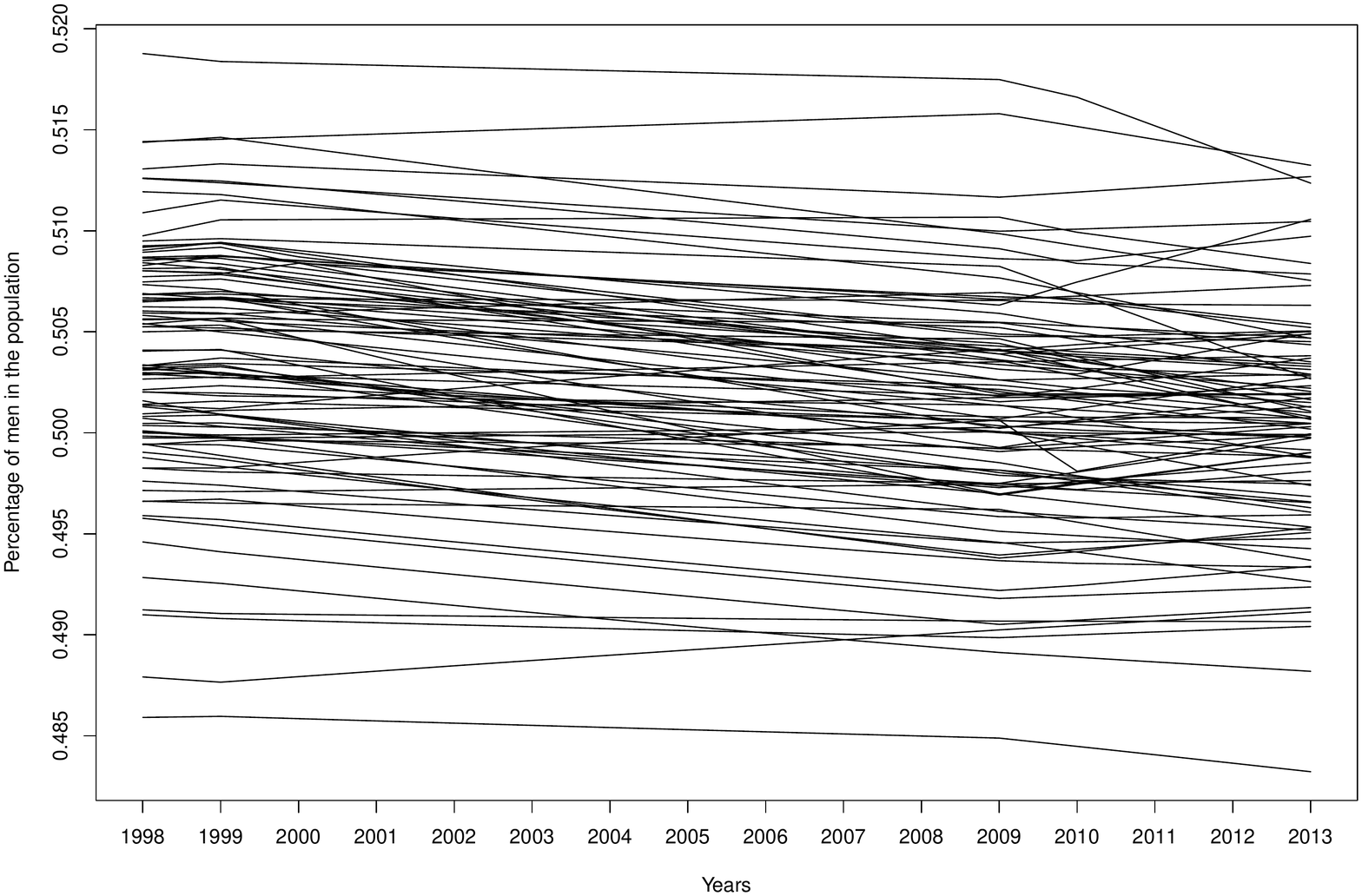} 
	\caption{Change over time in the proportion of men in the underlying population in France during the period from 1998 to 2013. Each curve corresponds to a \emph{d\'epartement}.}
	\label{Sex_graph}
\end{figure}


%
%

\end{document}